\documentclass[preprintnumbers,amsmath,amssymb,showpacs]{revtex4}
\usepackage{graphicx}
\usepackage{dcolumn}
\usepackage{bm}
\usepackage{booktabs}
\usepackage{CJK}
\usepackage{subfigure}
\usepackage{qcircuit}
\usepackage{color}
\usepackage{braket}
\usepackage{listings}
\usepackage{soul}
\usepackage{xcolor}
\usepackage{mathtools}
\usepackage{threeparttable} 
\usepackage{booktabs} 
\usepackage{multirow} 
\soulregister\cite7 
\soulregister\ref7 
\usepackage{color, soul} 
\setstcolor{yellow}

\begin{document}

\title{Symmetric joint measurement as a complement to the elegant joint measurement}

\author{Ying-Qiu He$^{1}$}
\author{Yu-Yan Zhang$^{1}$}
\author{Dong Ding$^{1,2}$}
\email{dingdong@ncist.edu.cn}
\author{Ting Gao$^{3}$}
\email{gaoting@hebtu.edu.cn}
\author{Feng-Li Yan$^{4}$}
\email{flyan@hebtu.edu.cn}

\affiliation {
$^1$ School of Science, University of Emergency Management, Beijing 101601, China\\
$^2$ Key Laboratory of Brain-Computer Interface Technology Application of the Ministry of Emergency Management, Beijing 101601, China\\
$^3$ School of Mathematical Sciences, Hebei Normal University, Shijiazhuang 050024, China\\
$^4$ College of Physics, Hebei Key Laboratory of Photophysics Research and Application,  Hebei Normal University, Shijiazhuang 050024, China\\
}
\date{\today}
\begin{abstract}
Traditional Bell state measurement (BSM) and product basis measurements (PBM) have been integral to nearly the entire development of quantum computing. Unlike the BSM and the PBM, a recently proposed two-qubit joint measurement called the elegant joint measurement (EJM) exhibits novel tetrahedral symmetry in its single-qubit reduced states.
In [Phys.Rev.Lett.126:220401], a parameterized two-qubit iso-entangled basis was proposed, with concurrence between 1/2 and 1, perfectly spanning the original EJM and conventional BSM.
We present a two-qubit symmetric joint measurement having concurrence from 0 to 1/2, which is complementary to [Phys.Rev.Lett.126:220401] and contains the PBM and the original EJM.
We investigate the symmetry of the current structure and its application in triangular networks. The results indicate that the reduction vectors of the current basis states exhibit rotational symmetry, rather than the aforementioned mirror symmetry; moreover, the output probability distributions of three parties in the network explicitly demonstrate the expected permutation symmetry.
Furthermore, we generalize the two-qubit symmetric joint measurement to the multiqubit systems with an even number of qubits.
\end{abstract}

\pacs{03.65.Ud; 03.67.-a; 03.67.Mn}


%
\maketitle

\section{Introduction}

Quantum measurement serves as a fundamental component in quantum information processing \cite{Quantum-entanglement,Entanglement-detection2009,IQIS2015}, functioning as a crucial bridge between quantum resources and classical information. The inherent uncertainty in measurement outcomes underpins the unique advantages of quantum computing, enabling phenomena such as quantum parallelism \cite{QST2017}.
A fundamental measurement operation in quantum information is single-qubit measurement \cite{NC2000}, typically performed as a projective measurement in a specific orthogonal basis (such as the computational basis $|0\rangle, |1\rangle$).
In contrast to single-qubit measurements, multiqubit joint measurements are capable of projecting  multiple qubits simultaneously. It reveals global system properties, such as network nonlocality \cite{Phys.Rev.Lett.104:170401(2010),PhysRevA.90.062109(2014),Gisin2014,PhysRevLett.120.140402(2018),FNN-PRL2022,Tavakoli-network2022} via correlated measurements typified by Bell state measurement (BSM).

In 2019, Gisin \cite{Gisin-EJM2019} introduced a novel two-qubit iso-entangled basis named elegant joint measurement (EJM). For the four basis states of EJM, each has a concurrence of $C=1/2$, and their single-qubit reduced states exhibit tetrahedral symmetry.
Subsequently, Tavakoli \emph{et al.} \cite{TGB-EJM2021} generalized the original EJM to a single-parameter version  that includes the traditional BSM.
More recently, He et al. \cite{He-PRA.111.012429(2025)} further extended the single-parameter EJM to a three-parameter EJM, parametrized by $z$, $\varphi$, and $\theta$. Ding et al. \cite{Ding-PRA-EJM2025} explored the extension of the two-qubit EJM to multiqubit systems based on the three-parameter EJM.
Concurrently, Del Santo \emph{et al.} \cite{PRR-EJM2024} offered a complete classification of two-qubit iso-entangled joint measurement bases, and Pauwels and Gisin \cite{arXiv:2509.01852} utilized the tetrahedral directions involved in the original EJM as basic units to explore the construction of multiqubit joint measurements.

Let us now return to the subject of the two-qubit symmetric joint measurements.
Consider the two-qubit state defined in Ref. \cite{He-PRA.111.012429(2025)}
\begin{eqnarray} \label{EJM-state}
|\Phi_{}\rangle=\frac{1}{2\sqrt{2}}  [(\sqrt{3}+\text{e}^{\text{i}\theta_{}})|m_{0},m_{1}\rangle
                                     +(\sqrt{3}-\text{e}^{\text{i}\theta_{}})|m_{1},m_{0}\rangle],
\end{eqnarray}
where real parameter $\theta_{} \in [0,\pi/2]$ and $\{|m_{0}\rangle, |m_{1}\rangle\}$ is a single-qubit orthogonal basis.
Based on this state, a family of two-qubit EJM bases can be constructed, containing the single-parameter EJM. Nevertheless, we note that nearly all existing parameterized EJM are confined to an achievable concurrence of $C = [1/2, 1]$.
We are interested in whether, beyond this restriction, there exist other possibilities for constructing symmetric joint measurements.
In this work, we redefine a two-qubit symmetric joint measurement basis. It serves as a complement to the EJMs, spanning the concurrence interval $C = [0, 1/2]$.
We investigate the symmetry of the reductions of the current basis states and elucidate its relationship to the original EJM.
As an application, we utilize this basis to detect network nonlocality in a triangular network, showing the permutation invariance of the resulting probability distributions. Finally, we extend the two-qubit symmetric joint measurement to multiqubit cases involving an even number of qubits.

\section{Two-qubit symmetric joint measurement basis}

\subsection{Definition of the basis states}

We consider a two-qubit symmetric joint measurement, which differs from the recently proposed single- or three-parameter EJMs whose concurrences lie in the interval $[1/2,1]$.
To achieve this, we first define two single-qubit states
\begin{eqnarray} \label{m-k-0}
|m^{}_{k,0}\rangle=\frac{1}{\sqrt{4+2\sqrt{2}}}
[(1+\text{e}^{-\text{i}\pi/4})|m_{k}\rangle + (1+\text{e}^{\text{i}\pi/4})|-m_{k}\rangle]
\end{eqnarray}
and
\begin{eqnarray} \label{m-k-1}
|m^{}_{k,1}\rangle=\frac{1}{\sqrt{4+2\sqrt{2}}}
[(1+\text{e}^{\text{i}\pi/4})|m_{k}\rangle  + (1+\text{e}^{-\text{i}\pi/4})|-m_{k}\rangle],
\end{eqnarray}
where
\begin{eqnarray} \label{m-k}
|\pm m^{}_{k}\rangle=\frac{1}{\sqrt{2}}   [\sqrt{1 \pm \cos(k\pi)}\text{e}^{-\text{i}\varphi_{k}/2}|0\rangle
                                       \pm \sqrt{1 \mp \cos(k\pi)}\text{e}^{\text{i}\varphi_{k}/2}|1\rangle], ~~k=0,1,2,3,
\end{eqnarray}
with $\varphi_{0}=\varphi_{}$, $\varphi_{1}=\varphi_{}+\pi/2$, $\varphi_{2}=\varphi_{}+\pi$, $\varphi_{3}=\varphi_{}-\pi/2$, and real parameter  $\varphi \in [-\pi, \pi]$.
Note that the states $|m_{k}\rangle$ and $|-m_{k}\rangle$ are orthogonal, $|m_{k,0}\rangle$ and $|m_{k,1}\rangle$ are non-orthogonal, i.e., $\langle m_{k}|-m_{k}\rangle=0$ and $\langle m_{k,0}|m_{k,1}\rangle=1/\sqrt{2}$.

We now define a set of two-qubit basis states
\begin{eqnarray} \label{2p-EJM-basis}
|\Phi_{k}\rangle=\frac{1}{2}
[(1+\text{e}^{\text{i}\theta_{}})|m_{k,0},m_{k,1}\rangle
+(1-\text{e}^{\text{i}\theta_{}})|m_{k,1},m_{k,0}\rangle], ~~k=0,1,2,3,
\end{eqnarray}
where real parameter $\theta \in [0, \pi/2]$.
One can observe that when $\theta=0$, these basis states are clearly product states $|\Phi_{k}(\theta=0)\rangle = |m_{k,0}, m_{k,1}\rangle$.
Substituting states (\ref{m-k-0}), (\ref{m-k-1}) and (\ref{m-k}), the present basis states (\ref{2p-EJM-basis}) in the computational basis can be expressed as
\begin{eqnarray} \label{simplified-basis}
|\Phi_{k}\rangle &=& \frac{1}{2}
(\text{e}^{-\text{i}\varphi_{k}}|00\rangle - r_{k,\theta}^{-}|01\rangle - r_{k,\theta}^{+}|10\rangle + \text{e}^{\text{i}\varphi_{k}}|11\rangle),
\end{eqnarray}
where
$r_{k,\theta}^{\pm}=[\cos(k\pi) \pm \text{i}\text{e}^{\text{i}\theta}]/\sqrt{2}$.
The orthogonality of these basis states is easily verified, i.e.
\begin{eqnarray} \label{}
\langle\Phi_{j}|\Phi_{k}\rangle = \frac{1}{4}
[1+2\cos (\varphi_{k}-\varphi_{j})+\cos(j\pi)\cos(k\pi)] = \delta_{jk}, ~~j, k=0,1,2,3.
\end{eqnarray}
In comparison with state (\ref{EJM-state}), this state differs in two respects: (i) the states
$|m_{k,0}\rangle$ and $|m_{k,1}\rangle$ are not orthogonal, and (ii) the coefficients differ from those of the original state.

\subsection{Symmetry}

\begin{figure}[h]
      \centering
      \includegraphics[width=3.9in]{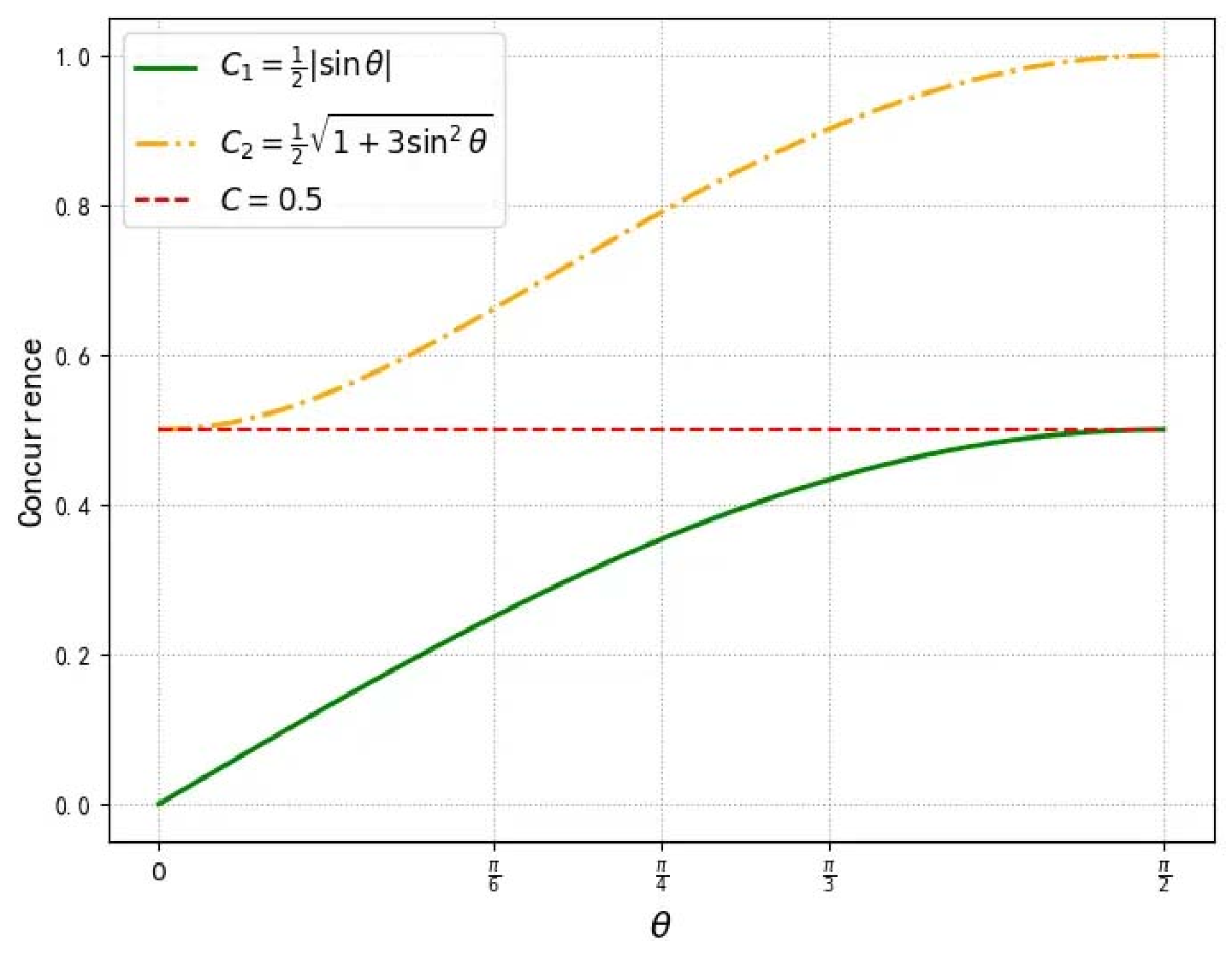}
      \caption{Comparison of concurrence for the current states (green solid line) and previous parameterized EJMs (orange dash-dot line). Both vary with the parameter $\theta$, distributed below and above $C=0.5$ (red dashed line, corresponding to the original EJM without parameters), respectively.}
\label{concurrence.eps}
  \end{figure}

Typical symmetries of the EJM basis states are reflected in iso-entanglement measure and the symmetry relations between their reduced states.
We first calculate the concurrence \cite{Quantum-entanglement} using
$C = \sqrt{2(1-\text{tr}\rho_{}^{2})}$, where $\rho_{}$ is a reduced state obtained by tracing out one qubit. A direct calculation gives
\begin{eqnarray}
C(|\Phi_{k}\rangle)  = \frac{1}{2}|\sin \theta|.
\end{eqnarray}
This is a non-trivial result with $C \in [0,1/2]$, which is distinct from the single- or three-parameter EJM cases, where the concurrence is given by $C = (1/2)\sqrt{1+3\sin^{2}\theta} \in [1/2,1]$.
Therefore, it can serve as a complement to the previous parameterized EJMs. The combination of the previous EJM basis states and our current basis states yields an entanglement measure that fully covers the $[0, 1]$ interval, as shown in Fig.\ref{concurrence.eps}.

\begin{figure}[h]
      \centering
      \includegraphics[width=3.0in]{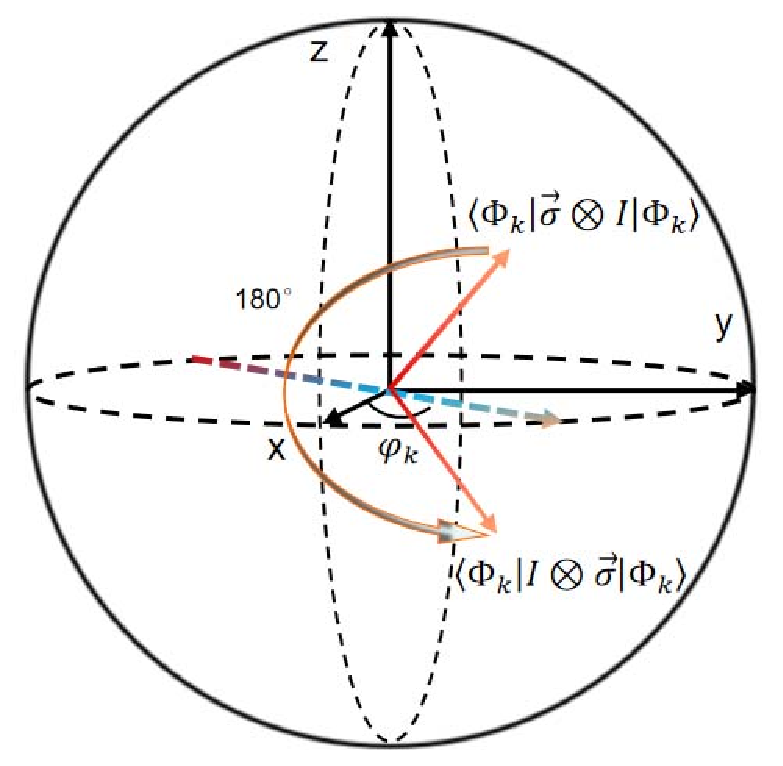}
      \caption{Rotational symmetry of the vectors corresponding to the reduced states $\langle \Phi_{k}|\vec{\sigma} \otimes I|\Phi_{k}\rangle$ and $\langle \Phi_{k}|I \otimes \vec{\sigma}|\Phi_{k}\rangle$. One vector can be transformed into the other by a 180-degree rotation about the axis $(\cos \varphi_{k}, \sin \varphi_{k}, 0)$.}
\label{reductions.eps}
  \end{figure}

We next see the reduced states of these basis states by calculate inner products $\langle \Phi_{k}|\vec{\sigma} \otimes I|\Phi_{k}\rangle$ and $\langle \Phi_{k}|I \otimes \vec{\sigma}|\Phi_{k}\rangle$. Here, $\vec{\sigma}=(\sigma_{x},\sigma_{y},\sigma_{z})$ is the vector of Pauli matrices, where $\sigma_{x},\sigma_{y},\sigma_{z}$ are three Pauli observables.
The calculation yields
\begin{eqnarray}
\langle \Phi_{k}|\vec{\sigma} \otimes I|\Phi_{k}\rangle
 &=& \frac{1}{\sqrt{2}} (-\cos (k\pi)\cos \varphi_{k}+ \cos \theta\sin \varphi_{k}, -\cos (k\pi)\sin \varphi_{k} -\cos \theta \cos \varphi_{k}, \frac{1}{\sqrt{2}} \cos (k\pi) \sin \theta)
\end{eqnarray}
and
\begin{eqnarray}
\langle \Phi_{k}|I \otimes \vec{\sigma}|\Phi_{k}\rangle
 &=&  \frac{1}{\sqrt{2}}(-\cos (k\pi) \cos \varphi_{k} - \cos \theta\sin \varphi_{k}, -\cos (k\pi)\sin \varphi_{k} + \cos \theta \cos \varphi_{k}, -\frac{1}{\sqrt{2}} \cos (k\pi) \sin \theta).
\end{eqnarray}
These two vectors are not mirror images of each other. Instead, they are connected through rotational symmetry. Specifically, either vector can be obtained by rotating the other by $180^\circ$ around the axis defined by the vector $(\cos \varphi_{k}, \sin \varphi_{k}, 0)$, i.e., a line in the $xy$-plane inclined at an angle $\varphi_{k}$ to the $x$-axis, as shown in Fig.\ref{reductions.eps}.
Moreover, substituting $k=0,1,2,3$ into the expressions above yields two tetrahedra with rotational symmetry.
Meanwhile, the constraint
\begin{eqnarray}
\sum_{k}\langle \Phi_{k}|\vec{\sigma} \otimes I|\Phi_{k}\rangle = \sum_{k} \langle \Phi_{k}|I \otimes \vec{\sigma}|\Phi_{k}\rangle =0
\end{eqnarray}
holds.

\subsection{Relation to the original EJM basis}

We next investigate the intrinsic connection between the current basis and the original EJM, which has a concurrence of $C=1/2$.
To do this, we set $\theta=\pi/2$, and the current basis states reduce to the following basis states
\begin{eqnarray} \label{}
|\Phi_{k}(\theta=\pi/2)\rangle &=& \frac{1}{2}
(\text{e}^{-\text{i}\varphi_{k}}|00\rangle
- r_{k}^{+}|01\rangle - r_{k}^{-}|10\rangle
+\text{e}^{ \text{i}\varphi_{k}}|11\rangle), ~~ k=0,1,2,3,
\end{eqnarray}
where
$r_{k}^{\pm} = [\cos (k\pi) \pm 1]/\sqrt{2}$.
Also, we rewrite the original EJM basis up to a global phase factor as
\begin{eqnarray} \label{}
|\Psi_{j}\rangle &=& \frac{1}{2}
(\text{e}^{ -\text{i}\phi_{j}}|00\rangle
- r_{j}^{+}|01\rangle - r_{j}^{-}|10\rangle
-\text{e}^{\text{i}\phi_{j}}|11\rangle), ~~ j=0,1,2,3,
\end{eqnarray}
where $\phi_{0} = 3\pi/4, \phi_{1} = -3\pi/4, \phi_{2} = -\pi/4, \phi_{3} = \pi/4$.
In order to compare these two bases, we calculate their inner product and have $\langle \Psi_{j}| \Phi_{k} (\theta=\pi/2) \rangle  = [1+\cos(j\pi)\cos(k\pi)+2\text{i}\sin (\phi_{j}-\varphi_{k})]/4$. Thus, for $k=j+1 ~(\text{mod} ~ 4)$ and $\phi_{j}-\varphi_{k}=n\pi$ ($n\in \mathbb{Z}$), the two basis states are orthogonal.
More specifically, by setting $\theta=\pi/2$ and $\varphi_{0} = \pi/4, \varphi_{1} = 3\pi/4, \varphi_{2} = -3\pi/4, \varphi_{3} = -\pi/4$,
we have $\langle \Psi_{0}| \Phi_{1} (\theta=\pi/2) \rangle = \langle \Psi_{1}| \Phi_{2} (\theta=\pi/2) = \langle \Psi_{2}| \Phi_{3} (\theta=\pi/2) = \langle \Psi_{3}| \Phi_{0} (\theta=\pi/2) =0$.
From this, it follows that the current basis is naturally reduced to the original EJM.

Therefore, analogous to the role of the single-parameter EJM as a bridge between the original EJM and BSM, the current basis is particularly notable for its ability to continuously span both the product basis and the original EJM.

\subsection{Quantum circuits}

\begin{figure}
\centerline{ \Qcircuit @C=1em @R=1em {
&\ctrl{1} &\gate{H} &\ctrl{1}  &\qw   &\gate{R_{x}({\pi}/{2}-2\varphi)} &\ctrl{1}   &\qw   &\gate{H}   &\meter \\
&\targ    &\qw      &\gate{R_{}({\pi}/{2}-\theta)} &\gate{X}  &\ctrl{-1}&\gate{S}  &\gate{X}  &\gate{H}   &\meter \\
}}
\vskip 0.55\baselineskip
\centerline{\footnotesize}
\caption{Quantum circuit for detecting the generalized three-parameter EJM.}
\label{quantum-circuit}
\vskip 0.55\baselineskip
\end{figure}
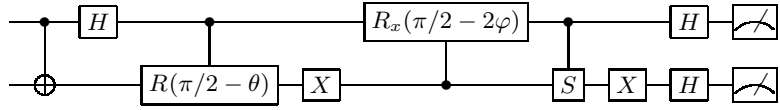

We here provide the quantum circuit for the current symmetric joint measurement, as shown in Fig.\ref{quantum-circuit}.
Our quantum circuit incorporates the conventional single-qubit gates $H=(\sigma_{x}+\sigma_{z})/\sqrt{2}$ and $X=\sigma_{x}$, as well as available two-qubit controlled gates: $\text{C}_{R({\pi}/{2}-\theta)}$, $\text{C}_{R_{x}({\pi}/{2}-2\varphi)}$, CNOT and the controlled-phase gate. Here, $R({\pi}/{2}-\theta)=\text{diag}[1, \text{e}^{\text{i}({\pi}/{2}-\theta)}]$, $S=R({\pi}/{2})$, and  $R_{x}({\pi}/{2}-2\varphi)=\text{e}^{-\text{i}({\pi}/{4}-\varphi)\sigma_{x}}$.
In this way, the four orthonormal basis states of the current symmetric joint measurement can be discriminated in the computational basis, as
\begin{eqnarray} \label{}
|\Phi_{0}\rangle \rightarrow |01\rangle, ~~ |\Phi_{1}\rangle \rightarrow -|11\rangle, ~~
|\Phi_{2}\rangle \rightarrow -|00\rangle, ~~ |\Phi_{3}\rangle \rightarrow |10\rangle.
\end{eqnarray}
Notably, with $\theta=\pi/2$ and $\varphi=\pi/4$, $R({\pi}/{2}-\theta)=R_{x}({\pi}/{2}-2\varphi)=I$, and the circuit recovers the original EJM up to local unitary operations.

\section{Application}

\begin{figure}[h]
      \centering
      \includegraphics[width=2.2in]{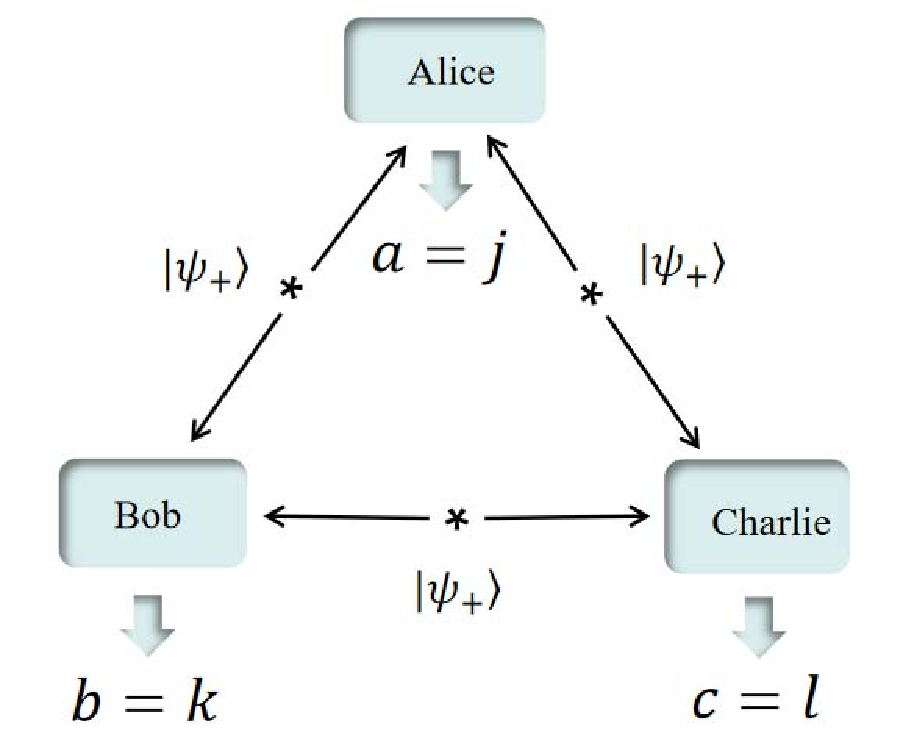}
      \caption{A triangular network of three parties connected by three independent sources. Each source creates a pair of Bell states $|\psi_{+}\rangle = (|01\rangle + |10\rangle)/\sqrt{2}$, which are distributed to neighboring parties. All parties perform the current symmetric joint measurements; the outputs for Alice, Bob, and Charlie are denoted by $a$, $b$, and $c$, respectively.}
\label{triangular-network.eps}
  \end{figure}

Here, we apply the current symmetric joint measurement basis to the triangular network, as shown in Fig.\ref{triangular-network.eps}. We assume that three independent sources each prepares a pair of Bell states $|\psi_{+}\rangle = (|01\rangle + |10\rangle)/\sqrt{2}$ and distributes them to the neighboring parties Alice and Bob, Bob and Charlie, and Charlie and Alice,
respectively.
The composite system reads
\begin{eqnarray} \label{}
|\Psi_{ABC}\rangle &=& |\psi_{+}\rangle_{A_{2}B_{1}} \otimes|\psi_{+}\rangle_{B_{2}C_{1}} \otimes|\psi_{+}\rangle_{C_{2}A_{1}}.
\end{eqnarray}
Then, the joint probability corresponding to network parties Alice, Bob, and Charlie resulting in
$|\Phi_{j}\rangle_{A_{1}A_{2}}$, $|\Phi_{k}\rangle_{B_{1}B_{2}}$, and $|\Phi_{l}\rangle_{C_{1}C_{2}}$, respectively, is
\begin{eqnarray}
p(a=j,b=k,c=l) = | \langle \Phi_{jkl} | \Psi_{ABC} \rangle |^2, ~~j, k, l =0,1,2,3,
\end{eqnarray}
where
$|\Phi_{jkl}\rangle = |\Phi_{j}\rangle_{A_{1}A_{2}}\otimes|\Phi_{k}\rangle_{B_{1}B_{2}}\otimes|\Phi_{l}\rangle_{C_{1}C_{2}}$.

Calculate
\begin{eqnarray}
\langle\Phi_{jkl}|\Psi_{ABC}\rangle &=& \frac{1}{32} \times \{
\text{e}^{-2\text{i}\theta}[\cos(j\pi)+\cos(k\pi)+\cos(l\pi)]
+2\text{e}^{-\text{i}\theta}[\sin(\phi_j-\phi_k)+\sin(\phi_k-\phi_l)+\sin(\phi_l-\phi_j)]\nonumber\\
&&-2[\cos(j\pi)\cos(\phi_k-\phi_l)
+\cos(k\pi)\cos(\phi_l-\phi_j)
+\cos(l\pi)\cos(\phi_j-\phi_k)] \nonumber\\
&&
-\cos(j\pi)\cos(k\pi)\cos(l\pi)
\}.
\end{eqnarray}
Then we have
\begin{eqnarray}
p(a=j,b=k,c=l) =
\begin{dcases}
\frac{4 + 21 \sin^2\theta}{256}, ~ \text{if} ~ j=k=l, &  \\
\frac{4+\sin^2\theta}{256}, ~~~~~ \text{if}~ j \neq k \neq l \neq j, &  \\
\frac{4-3\sin^2\theta}{256}, ~~~ \text{else}, &
\end{dcases}
\end{eqnarray}
satisfying the normalization condition
\begin{eqnarray}
\frac{1}{256}[(4 + 21 \sin^2\theta) \times 4 + (4 + \sin^2\theta)\times 24 + (4-3\sin^2\theta) \times 36] = 1.
\end{eqnarray}

It is observed that the output probabilities of the measurement results vary with the real parameter $\theta$, always maintaining permutation invariance. In particular, for  $\theta=\pi/2$, the probabilities simplify to those of the original EJM case:
$p(a=b=c=j) = {25}/{256}$, $p(a=j,b=k,c=l) = {5}/{256}$ (where $j \neq k \neq l \neq j$), and $p(a=b=j,c=l) = {1}/{256}$ for $j \neq l$, as well as their cyclic permutations. In contrast, for $\theta=0$, the probabilities become uniform with $p=1/64$, corresponding to all parties performing measurements in the product basis.

Consider a trilocal model \cite{Gisin-EJM2019} within the triangular network, where the maximal three-particle trilocal correlation is
$p_{\text{max}}(a=b=c) = {61}/{256}$. Once this bound is surpassed by the quantum system, the nonlocality of the quantum network is revealed.
A direct calculation results in
\begin{flalign}
p(a=b=c) = \frac{4 + 21 \sin^2\theta}{64};
\end{flalign}
notably, quantum nonlocality appears immediately for $\arcsin (\sqrt{15/28}) < \theta \leq \pi/2$.

\section{Multiqubit symmetric joint measurement basis}

In this section, we extend the current symmetric joint measurement basis to the even-$n$ case.
Concretely, for even $n$, an $n$-qubit symmetric joint measurement basis is defined as
\begin{eqnarray} \label{n-qubit-basis}
|\Phi_{k_1 \cdots k_{n/2}}^{}\rangle &=& \frac{1}{2}
   [(1+\text{e}^{\text{i}\theta_{}})|m_{k_1,0},m_{k_1,1},  \cdots ,m_{k_{n/2},0},m_{k_{n/2},1}\rangle
  + (1-\text{e}^{\text{i}\theta_{}})|m_{k_1,1},m_{k_1,0},\cdots ,m_{k_{n/2},1},m_{k_{n/2},0}\rangle],
\end{eqnarray}
where $k_1, \cdots, k_{n/2}=0,1,2,3$.
To verify the orthonormality of the basis state, we calculate inner product
\begin{eqnarray}
\langle \Phi_{j_1 \cdots j_{n/2}}|\Phi_{k_1 \cdots k_{n/2}}\rangle &=& \frac{1}{4}
(|1+\text{e}^{\text{i}\theta}|^{2}
\langle m_{j_1,0}|m_{k_1,0}\rangle \langle m_{j_1,1}|m_{k_1,1}\rangle
\cdots \langle m_{j_{n/2},0}|m_{k_{n/2},0}\rangle \langle m_{j_{n/2},1}|m_{k_{n/2},1}\rangle \nonumber \\ &&
+|1-\text{e}^{\text{i}\theta}|^{2} \langle m_{j_1,1}|m_{k_1,1}\rangle \langle m_{j_1,0}|m_{k_1,0}\rangle
\cdots \langle m_{j_{n/2},1}|m_{k_{n/2},1}\rangle \langle m_{j_{n/2},0}|m_{k_{n/2},0}\rangle).
\end{eqnarray}

To simplify the calculation, we here define two sets of orthonormal bases
\begin{eqnarray} \label{m-0}
|m^{\pm}_{0}\rangle=\frac{1}{\sqrt{4+2\sqrt{2}}}
[(1+\text{e}^{-\text{i}\pi/4})\text{e}^{-\text{i}\varphi_{}/2}|0\rangle \pm (1+\text{e}^{\text{i}\pi/4})\text{e}^{\text{i}\varphi_{}/2}|1\rangle],
\end{eqnarray}
and
\begin{eqnarray} \label{m-1}
|m^{\pm}_{1}\rangle=\frac{1}{\sqrt{4+2\sqrt{2}}}
[(1+\text{e}^{\text{i}\pi/4})\text{e}^{-\text{i}\varphi_{}/2}|0\rangle \pm (1+\text{e}^{-\text{i}\pi/4})\text{e}^{\text{i}\varphi_{}/2}|1\rangle],
\end{eqnarray}
satisfying $\langle m^{+}_{0}|m^{-}_{0}\rangle = \langle m^{+}_{1}|m^{-}_{1}\rangle =0$.
Accordingly, $|m_{k,0}\rangle$ and  $|m_{k,1}\rangle$ can be written as
\begin{eqnarray} \label{}
|m_{0,0}\rangle = |m^{-}_{0}\rangle,      ~~     |m_{1,0}\rangle = |m^{+}_{0}\rangle, ~~
|m_{2,0}\rangle = -\text{i} |m^{+}_{0}\rangle,~~ |m_{3,0}\rangle = \text{i} |m^{-}_{0}\rangle,
\end{eqnarray}
and
\begin{eqnarray} \label{}
|m_{0,1}\rangle = |m^{-}_{1}\rangle,          ~~ |m_{1,1}\rangle = -\text{i} |m^{-}_{1}\rangle, ~~
|m_{2,1}\rangle = -\text{i} |m^{+}_{1}\rangle, ~~|m_{3,1}\rangle = |m^{+}_{1}\rangle.
\end{eqnarray}
From this, we can immediately derive  $\langle m_{j_1,0}|m_{k_1,0}\rangle \langle m_{j_1,1}|m_{k_1,1}\rangle = \delta_{j_1k_1}$,
and thus, we have
\begin{eqnarray}
\langle \Phi_{j_1 \cdots j_{n/2}}|\Phi_{k_1 \cdots k_{n/2}}\rangle &=& \delta_{j_1k_1} \cdots \delta_{j_{n/2}k_{n/2}},
~~ j_{1}, \cdots,  j_{n/2},k_{1}, \cdots, k_{n/2}=0,1,2,3.
\end{eqnarray}
Therefore, the states (\ref{n-qubit-basis}) constitute an $n$-qubit orthogonal basis, which comprise product bases when $\theta=0$.

The symmetry of these basis states is revealed by calculating the inner product
$\langle \Phi_{k_1 \cdots k_{n/2}}|(\vec{\sigma} \otimes I \otimes \cdots \otimes I)_{\text{perm}}|\Phi_{k_1 \cdots k_{n/2}}\rangle$, where $(.)_{\text{perm}}$ stands for all permutations of the operators.
The results of the inner products are given by the following expression
\begin{eqnarray}
I_{k_i}^{\pm}&=& \frac{1}{2}
[(\langle m_{k_i,0}|\vec{\sigma}| m_{k_i,0}\rangle + \langle m_{k_i,1}|\vec{\sigma}| m_{k_i,1}\rangle)
\pm \cos \theta
(\langle m_{k_i,0}|\vec{\sigma}| m_{k_i,0}\rangle - \langle m_{k_i,1}|\vec{\sigma}| m_{k_i,1}\rangle)
\nonumber \\ &&
\mp 2^{\frac{1-n}{2}} \times \text{i}\sin \theta
(\langle m_{k_i,0}|\vec{\sigma}| m_{k_i,1}\rangle - \langle m_{k_i,1}|\vec{\sigma}| m_{k_i,0}\rangle)
], ~~ i=1,2, \cdots,  {n}/{2}.
\end{eqnarray}
A not very complicated calculation yields
\begin{eqnarray}
I_{k_i}^{\pm}&=& \frac{1}{\sqrt{2}}
(-\cos (k_i\pi)\cos \varphi_{k_i} \pm \cos \theta \sin \varphi_{k_i},
-\cos (k_i\pi)\sin \varphi_{k_i} \mp \cos \theta \cos \varphi_{k_i},
\pm 2^{\frac{1-n}{2}} \cos (k_i\pi) \sin \theta).
\end{eqnarray}
Therefore, the rotational symmetry of these reductions is preserved.

For clarity, taking $n=2$ and setting $\theta=\pi/2$ and $\varphi_{0} = \pi/4, \varphi_{1} = 3\pi/4, \varphi_{2} = -3\pi/4, \varphi_{3} = -\pi/4$, we have
$I_{0}^{\pm}=(1/2)(-1,-1,\pm 1)$, $I_{1}^{\pm}=(1/2)(-1,1,\mp 1)$, $I_{2}^{\pm}=(1/2)(1,1,\pm 1)$, and $I_{3}^{\pm}=(1/2)(1,-1,\mp 1)$.
That is to say, $I_{k_i}^{+}$ and $I_{k_i}^{-}$  ($k_i=0,1,2,3$) form two mirror-image regular tetrahedra with radius $\sqrt{3}/2$, corresponding explicitly to the original EJM.

\section{Discussion and summary}

In summary, we have proposed a two-qubit symmetric joint measurement, which differs from the previous single- or three-parameter EJMs whose concurrences lie in the interval $[1/2, 1]$.
Although its expression is quite similar to the previous single-parameter EJM, its concurrence falls within [0, 1/2], making it an effective complement to the previous EJMs.
By introducing a real parameter $\theta$, the single-parameter EJM incorporates BSM and the original EJM; likewise, the current basis bridges product bases ($\theta=0$) and the original EJM ($\theta=\pi/2$).

We have investigated the symmetry of the current basis states by analyzing the relationships between their single-qubit reductions.
Distinct from the mirror symmetry of the reduced states in the previous EJMs, our derived reduced states feature rotational symmetry. In particular cases, such as $\theta=\pi/2$, this symmetry can degenerate to mirror symmetry.
In applications, we apply the current symmetric joint measurement to a triangular network with three independent sources. As a result, this structure not only can reveal network nonlocality for $\arcsin (\sqrt{15/28}) < \theta \leq \pi/2$, but the output probability distributions of the three parties also exhibit the expected permutation symmetry.
Moreover, we have extended the two-qubit symmetric joint measurement to multiqubit systems consisting of an even number of qubits.

Several aspects of this work remain to be further improved.
(1) The proposed two-qubit basis includes product basis and the original EJM. It covers the same concurrence interval [0,1/2] as the `elegant family' in Ref. \cite{PRR-EJM2024}, yet the specific relationship between the two bases remains unclear.
(2) The tetrahedral rotational symmetry of the single-qubit reduced states found in this work is conceptually more general than the previous mirror symmetry. It remains an open question whether this rotational symmetry should be a primary consideration in future extensive studies of symmetry measurements.
(3) In terms of applications, beyond the network nonlocality that has already been experimentally realized \cite{BGT2021IBM,GuoPRL-EJM2022}, it is worth asking whether there exist other applications capable of surpassing BSM. After all, in quantum teleportation using EJM \cite{QT-EJM2024}, aside from offering a wider range of measurement settings, no more exciting results have been obtained.
(4) We generalize two-qubit symmetric joint measurements to the multiqubit systems, specifically focusing on the even-number cases. While the strategy in Ref. \cite{Ding-PRA-EJM2025} allows for the extension to odd-number cases, achieving a generalization for an arbitrary number of qubits or higher-dimensional systems \cite{Quantum.5.442} remains an open problem.
These parameterized structures extend conceptual frameworks and enrich symmetric joint measurements, while offering new perspectives for future applications.

\begin{acknowledgements}
We thank J. Czartowski and K. \.{Z}yczkowski for bringing Ref. \cite{Quantum.5.442} to our attention.
This work was supported by
the Hebei Science and Technology Program Foundation of China under Grant No. 246Z0902G.
\end{acknowledgements}

\end{document}